\documentclass[12pt,preprint]{aastex}

\usepackage{emulateapj5}

\usepackage{amsmath,amsthm}

\usepackage{graphicx}

\def\go{\mathrel{\raise.3ex\hbox{$>$}\mkern-14mu
             \lower0.6ex\hbox{$\sim$}}}

\def\lo{\mathrel{\raise.3ex\hbox{$<$}\mkern-14mu
             \lower0.6ex\hbox{$\sim$}}}

\newenvironment{inlinefigure}{
\medskip
\def\@captype{figure}
\noindent\begin{minipage}{0.999\linewidth}\begin{center}}
{\end{center}\end{minipage}\medskip}

\begin{document}

\title{The Light Curve and Internal Magnetic Field of the Mode-Switching Pulsar PSR B0943+10}

\author{Natalia I. Storch\altaffilmark1, Wynn C.G. Ho\altaffilmark2, Dong Lai\altaffilmark1, Slavko Bogdanov\altaffilmark3, and Craig O. Heinke\altaffilmark4}
\altaffiltext1{Center for Space Research, Department of Astronomy, Cornell University, Ithaca, NY 14853}
\altaffiltext2{Mathematical Sciences and STAG Research Centre, University of Southampton, Southampton, SO17 1BJ, United Kingdom}
\altaffiltext3{Columbia Astrophysics Laboratory, Columbia University, 550 West 120th Street, New York, NY 10027}
\altaffiltext4{Department of Physics, University of Alberta, CCIS 4-181, Edmonton, AB, T6G 2E1, Canada}

\begin{abstract}

A number of radio pulsars exhibit intriguing mode-switching behavior.
Recent observations of PSR~B0943+10 revealed correlated radio and
X-ray mode switches, providing a new avenue for understanding this
class of objects.  The large X-ray pulse fraction observed during the
radio quiet phase (Q mode) was previously interpreted as a result of
changing obscuration of X-rays by dense magnetosphere plasma.  We show
that the
large X-ray pulse fraction can be explained by including the beaming effect
 of a magnetic atmosphere, while remaining consistent with the
dipole field geometry constrained by radio observations. We also
explore a more extreme magnetic field configuration, where a magnetic
dipole displaced from the center of the star produces two magnetic
polar caps of different sizes and magnetic field strengths. These models are
currently consistent with data in radio and X-rays and can be tested
or constrained by future X-ray observations.

\end{abstract}


\section{Introduction}

It has long been known that many radio pulsars exhibit
variabilities in their pulse shapes and intensities, on a wide variety
of timescales. These include phenomena such as subpulse
drifts, nulls and bursts, and long- and short-term on/off quasiperiodic
mode-switching (see Cordes 2013 for a review).  Until recently, these
variabilities were thought to be due to changes in the radio emission
process, not tied to the global properties of the pulsar
magnetosphere and high energy emission.  However, Kramer et al.~(2006)
discovered that PSR~B1931+24 shows large changes in its spin-down
rate, correlated with its mode switching, suggesting that the
switching is strongly tied to the global energetics of the
pulsar. Later, PSR~J1841-0500 (Camilo et al. 2012) and PSR~J1832+0029
(Lorimer et al. 2012) were also shown to have similar behavior.  A
number of other pulsars display smaller changes in spin-down rates
which are correlated with changes in their average pulse profiles (Lyne et
al.~2010). These may correspond to a less extreme version of  ``mode-switching''.

This connection between radio mode-switching and global magnetosphere change
was further strengthened by the simultaneous X-ray and radio
observations of PSR~B0943+10 (Hermsen et al.~2013; hereafter H13). 
This is a relatively old pulsar which has been known
for a long time to transition from a bright, highly organized radio ``on''
(B) state to a quieter, more chaotic radio ``off'' (Q) state on timescales
of a few hours (Rankin \& Suleymanova
2006). In addition, its radio emission exhibits subpulse drift, which
was shown by Deshpande \& Rankin (2001) to likely come from $20$
`sub-beams' arranged in a conical structure. Deshpande \& Rankin
(2001) also constrained the emission geometry of the radio
beam, demonstrating that the line of sight, rotation axis, and
magnetic axis are all nearly parallel, with a misalignment of only
$\sim\!10^{\circ}$. H13 found that the
radio mode-switching of PSR~B0943+10 is correlated with a similar
switching in its X-ray emission: during the B mode the X-ray emission
is steady and non-pulsed, while during the Q mode the X-ray flux
increases by a factor of $\sim\!2$ and becomes highly pulsed.

As PSR~B0943+10 is a paragon of mode-switching pulsars, understanding
its behavior may provide important insight into the physical origin of
pulsar mode-switching.  In this paper, we discuss three puzzling
aspects of PSR~B0943+10's correlated X-ray and radio emission (section
2) and demonstrate that two of them can be resolved through accurate
light curve modeling that accounts for the effect of a strong magnetic
field (section 3). 
We consider an alternative, ``displaced dipole'' magnetic field geometry,
and show that it 
provides a satisfactory explanation for all three puzzles (section
4). In section 5 we summarize our findings, discuss their
implications, and propose concrete methods by which they can be
refuted or confirmed.


\section{Observational Puzzles of PSR~B0943+10} 

Of the several puzzles that PSR~B0943+10 presents, the most important is, 
undoubtedly, the question of \textit{how} and \textit{why} the
mode-switching occurs. The physical mechanism for mode change is unknown
(see Cordes 2013) and we do not attempt to tackle it directly in this paper.
Instead, we address several ``smaller'' puzzles that
arose with the recent simultaneous X-ray and radio observations of
this pulsar (H13).

First, the X-ray observation showed that the pulsar exhibits a steady
unpulsed emission during the pulsar's radio-bright (B) mode, and a
larger, strongly pulsed emission during the radio-quiet mode (Q). The
emission during the radio-bright mode is suggested to be non-thermal in
nature, while the added pulsed component in the radio-quiet mode is
consistent with being thermal (H13).
However, the somewhat limited data at present is well fitted by either 
a blackbody \textit{or} a power law in both the Q and the B modes 
(Mereghetti et al. 2013; hereafter M13). 

Assuming the unpulsed radiation does not cease
when the pulsed component is present, we infer that the additional thermal
emission must have nearly $100\%$ pulse fraction. However, since
Deshpande \& Rankin (2001) showed that all three axes of the pulsar
(the observer line of sight, the rotation axis, and the magnetic
dipole axis; see Fig.~1) are aligned within $\sim\! 10^{\circ}$, at first glance it
appears unlikely that such a high pulse fraction can be achieved via
the standard polar cap hot spot. 
In section 3 we demonstrate that, in fact, this problem is remedied by
the inclusion of magnetic beaming.

H13 proposed that the strong pulsation may be
explained by periodic obscuration of the X-ray emission via electron
scattering in the magnetosphere. 
The resonant scattering cross-section $\sigma$ for photons with
frequency $\omega$ can be written as (e.g., Canuto et al.~1971)
\begin{equation}
\sigma = {4\pi^2e^2\over m_e c} |\varepsilon_-|^2 
\delta(\omega-\omega_c),
\end{equation}
where $e$ and $m_e$ are the electron charge and mass, respectively,
and $\omega_c\equiv eB(r)/m_ec$ is the cyclotron frequency at a
distance $r$ from the neutron star (NS), and $\varepsilon_-$ specifies
the circular polarization (around the magnetic field) of the photon.
We characterize the magnetosphere particle (electron or positron) 
number density $n(r)$ in terms of the Goldreich-Julian density,
i.e., $n(r)\simeq \lambda_{\rm GJ} \Omega B(r)/(2\pi ec)$, with
$\lambda_{\rm GJ}$ the multiplicity factor. Using $B(r)\simeq B_\star 
(R_\star/r)^3$ and $|\varepsilon_-|\simeq 1$, we find that the photon optical 
depth $\tau=\int_{R_\star}^\infty n\sigma dr$ is 
\begin{equation}
\tau \simeq 10^{-3}\,\lambda_{\rm GJ}\!\left({P\over 1~\mathrm{sec}}
\right)^{-1}\!\left({R_\star\over 10~\mathrm{km}}\right)
\left({B_\star\over 10^{12}~\mathrm{G}}\right)^{1\over 3}
\!\left({E\over 1~\mathrm{keV}}\right)^{-{1\over 3}},
\label{eq:tau}\end{equation}
where $P$, $R_\star$, and $B_\star$ are the period, radius, and
surface magnetic field of the pulsar, and $E=\hbar\omega$ is photon energy.
For the range of photon energies of $0.2-2$~keV in which the X-ray
pulsation is observed, in order for sufficient obscuration to occur
($\tau \go 1$), the density of the plasma must be about $1000$ times
the Goldreich-Julian density. 
While theoretically possible, it is unclear such a large
particle density can be achieved in the closed magnetosphere region.
Observationally, many isolated NSs (including both active 
radio pulsars and pure thermally emitting sources) have detectable 
X-ray emission from the whole stellar surface (e.g., Kaspi et al. 2004;
Zavlin 2007), indicating that obscuration of surface 
X-rays by the magnetosphere is not significant.

The second puzzle of PSR~B0943+10 is the presence of ``drifting
subpulses'' in its radio emission (Deshpande \& Rankin 2001).
These drifting subpulses are thought to be a consequence of 
${\bf E}\times{\bf B}$ drifting of a pattern of ``sparks'' in the polar cap
accelerator of the NS (Ruderman \& Sutherland 1975; Ruderman \& Gil 2006).
This implies the presence of a ``vacuum gap'' - a region directly over
the polar cap in which the pulsar magnetosphere has pulled away from
the surface of the NS, leaving a gap in which no plasma is
present\footnote{It is well known that a full vacuum gap potential would lead to
too fast drifting subpulses, so a partially screened gap may be required
(e.g., Gil, Melikidze \& Geppert 2003; van Leeuwen \& Timokhin 2012). See
also Jones (2014) for an alternative model for drifting subpulses in pulsars.}. 
~This gap~ then~ has~ a strong voltage

\begin{inlinefigure}
\scalebox{0.65}{\rotatebox{0}{\includegraphics{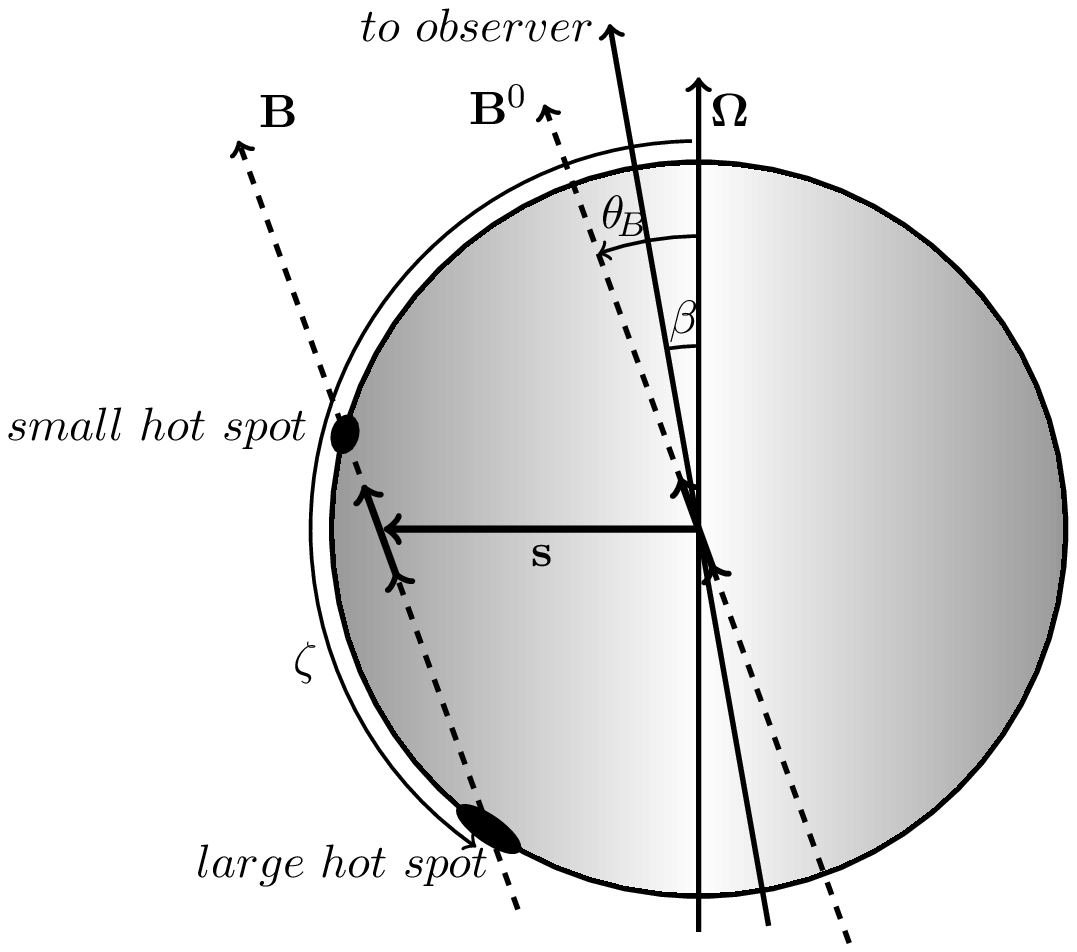}}}
\figcaption{A schematic picture of the emission geometry of
  PSR~B0943+10, and the proposed displaced dipole model of section 4. The
  misalignment of axes is exaggerated for clarity. The angles
  $(\beta,\theta_B)$ have been constrained by Deshpande \& Rankin
  (2001) to be in the range between
  $(7^\circ,11^\circ)$ and $(9^\circ,15^\circ)$.}
\end{inlinefigure}

\noindent drop across it, acting as
an accelerator for charged particles and causing them to emit
high-energy radiation, which later decays into highly bunched radio
emission via a pair-production cascade a few tens of NS radii away
from the surface. The presence of this gap is usually predicated upon
the inability of ions to become dissociated from the NS
surface. According to the electronic structure calculations of
condensed matter in strong magnetic fields by
Medin \& Lai (2006), for an iron NS surface, a high surface magnetic
field ($\go 10^{14}$~G, see Fig.~7 of Medin \& Lai 2007) is required
to keep the ions bound to the surface. The required field strength is
even larger if the condensed surface is made of lighter elements (C or
He).  On the other hand, the surface dipole magnetic field inferred
from the $P$-$\dot{P}$ measurement of PSR~B0943+10 is a modest
$2\times 10^{12}$~G, which is certainly insufficient for a gap to
form. This implies that perhaps the local surface magnetic field at
the polar cap of the NS is much higher than the dipole field inferred
from $P$-$\dot{P}$.

Finally, the fitted emission area of the thermal hot spot (M13) is found 
to be much smaller ($20-30$~m) than the canonical polar cap
size, $R_{\rm cap}=R_\star\sqrt{R_\star\Omega/c}=140$~m.
In sections 3 and 4 we demonstrate that, with proper atmospheric modeling,
the emission area can be consistent with the polar cap size.

\section{Light curves and the canonical emission geometry}

H13 found that the pulsed component of the X-ray emission
during the Q mode is comprised of a
thermal blackbody that has a pulse fraction close to $100\%$.
As discussed in Sec.~2, this poses a 
challenge to models of emission from a standard polar cap hot spot.
However, the determination of the pulse fraction is uncertain.
H13 showed the pulse fraction (before subtracting 
the unpulsed flux) to be $10\%\pm 15\%$ at 0.2-0.5~keV, and thus the pulse
fraction at low energies can be low and is ~even consistent with zero.

\begin{inlinefigure}
\scalebox{0.4}{\rotatebox{0}{\includegraphics{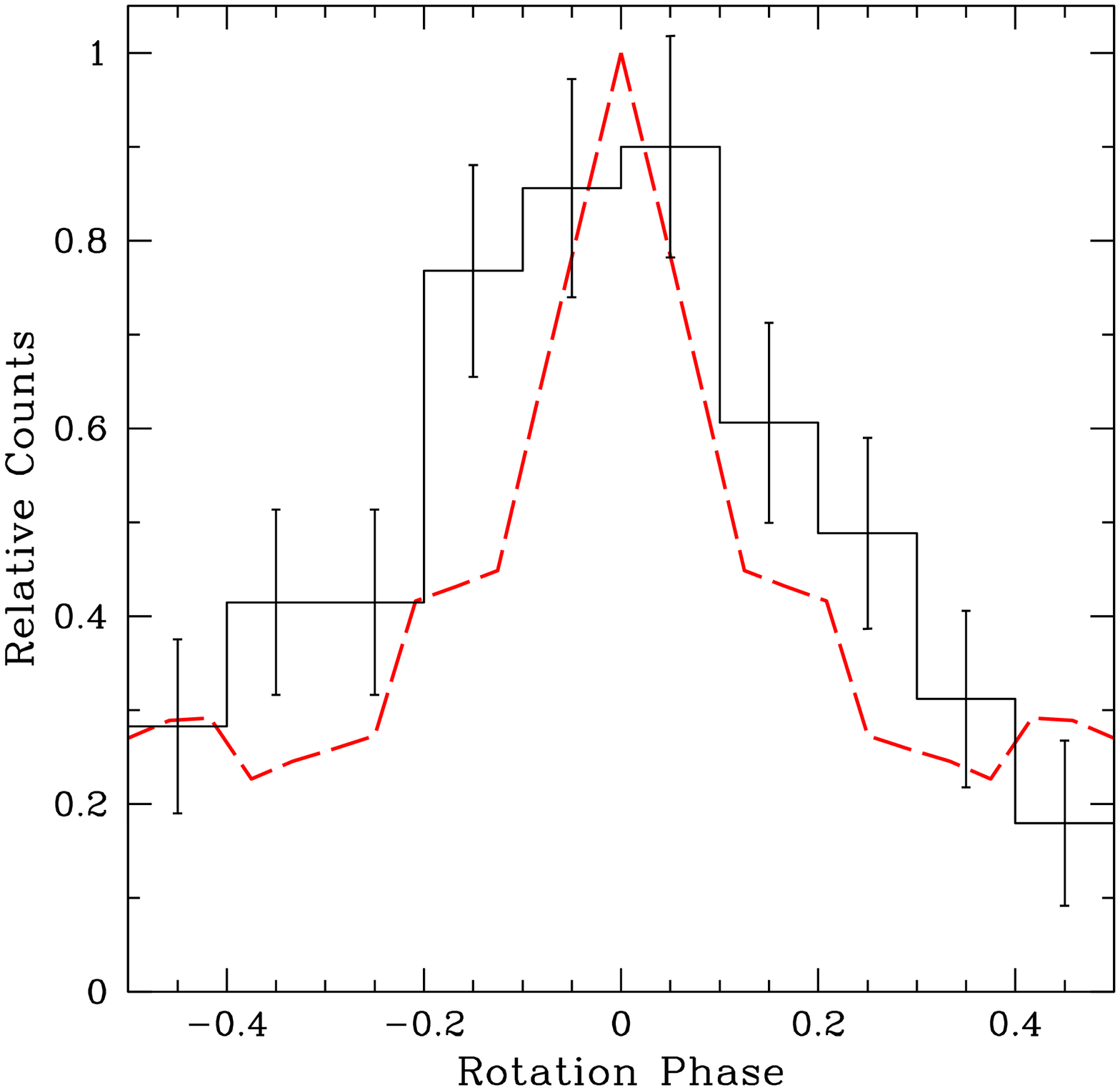}}}
\figcaption{Q mode X-ray pulse profile of PSR~B0943+10.
The histogram is the observed (0.15--12 keV) pulse profile taken from Fig.~3 of
M13, where we subtracted the background counts and
normalised by the maximum. Errors are $\approx 1\sigma$.
The dashed curve is the result of our atmosphere model with
$B=2\times 10^{12}\mbox{ G}$,
$M=1.2\,M_{\odot}$ and $R_\star=12$~km ($1+z=1.19$),
$R_{\mathrm{em}}\approx 85\mbox{ m}$,
$T_{\mathrm{eff}}=1.5\times 10^6\mbox{ K}$, and
$(\beta,\theta_B)=(8^\circ,14^\circ)$.}
\end{inlinefigure}

\noindent This is supported by the analyses of M13 who found
(by fitting the pulse profiles with a constant plus sinusoid) pulse
fractions of $66\%\pm 13\%$, $56\%\pm 8\%$, $53\%\pm 13\%$ at 0.15-0.6~keV,
0.6-1.3~keV, and 1.3-12~keV, respectively.  We also see from Fig.~3 of
M13 that the [(max-min)/total] pulse fractions are
$\approx 40-100\%$, $\approx 45-75\%$, and $\approx 35-100\%$, respectively,
accounting approximately for the $1\sigma$ errors in the pulsed and background
components (see Figure~3).

Since these pulse fractions may be less than 100\%, we now attempt to obtain
standard polar cap hot spot solutions that are consistent with the observed
spectrum, pulse fraction, and pulse profile of PSR~B0943+10.
We employ a magnetized partially ionized hydrogen atmosphere model
(Ho et al.~2008) to generate X-ray spectra and pulse profiles with a
procedure that accounts for relativistic effects (Ho~2007).
We take a magnetic field of $2\times 10^{12}\mbox{ G}$ 
(the dipole field inferred from $P$ and $\dot P$)
that is directed perpendicular to the surface normal.
We denote by $\beta$ the angle
between the line of sight and the rotation axis, and by $\theta_B$ the
angle between the magnetic and rotation axes (see Figure 1).
We assume $\beta=7-9^\circ$ and $\theta_B=11-15^\circ$, as constrained by 
Deshpande \& Rankin (2001),
and NS mass and radius $M=1.2\,M_{\odot}$ and $R_\star=12$~km, so that
the gravitational redshift is $1+z=1.19$.
Whether we include
or neglect an antipodal hot spot has no effect on our results since this second hot
spot is not visible for the assumed viewing geometry and gravitational redshift.
Here we describe our approximate fits, leaving detailed fits for future work
with higher-quality data.

We find that an atmosphere with
$T_{\mathrm{eff}}\approx (1.4-1.5)\times 10^6\mbox{ K}$ and
emission radius $R_{\mathrm{em}}\approx 85\mbox{ m}$ matches qualitatively the
radio-quiet X-ray spectrum of PSR~B0943+43 from H13.
Previous blackbody fits find $T_{\mathrm{BB}}=3\times 10^6\mbox{ K}$ and
$R_{\mathrm{BB}}\approx 20-30\mbox{ m}$
(H13, M13);
note that the lower temperature and larger emitting area are typical
characte-

\begin{inlinefigure}
\scalebox{0.4}{\rotatebox{0}{\includegraphics{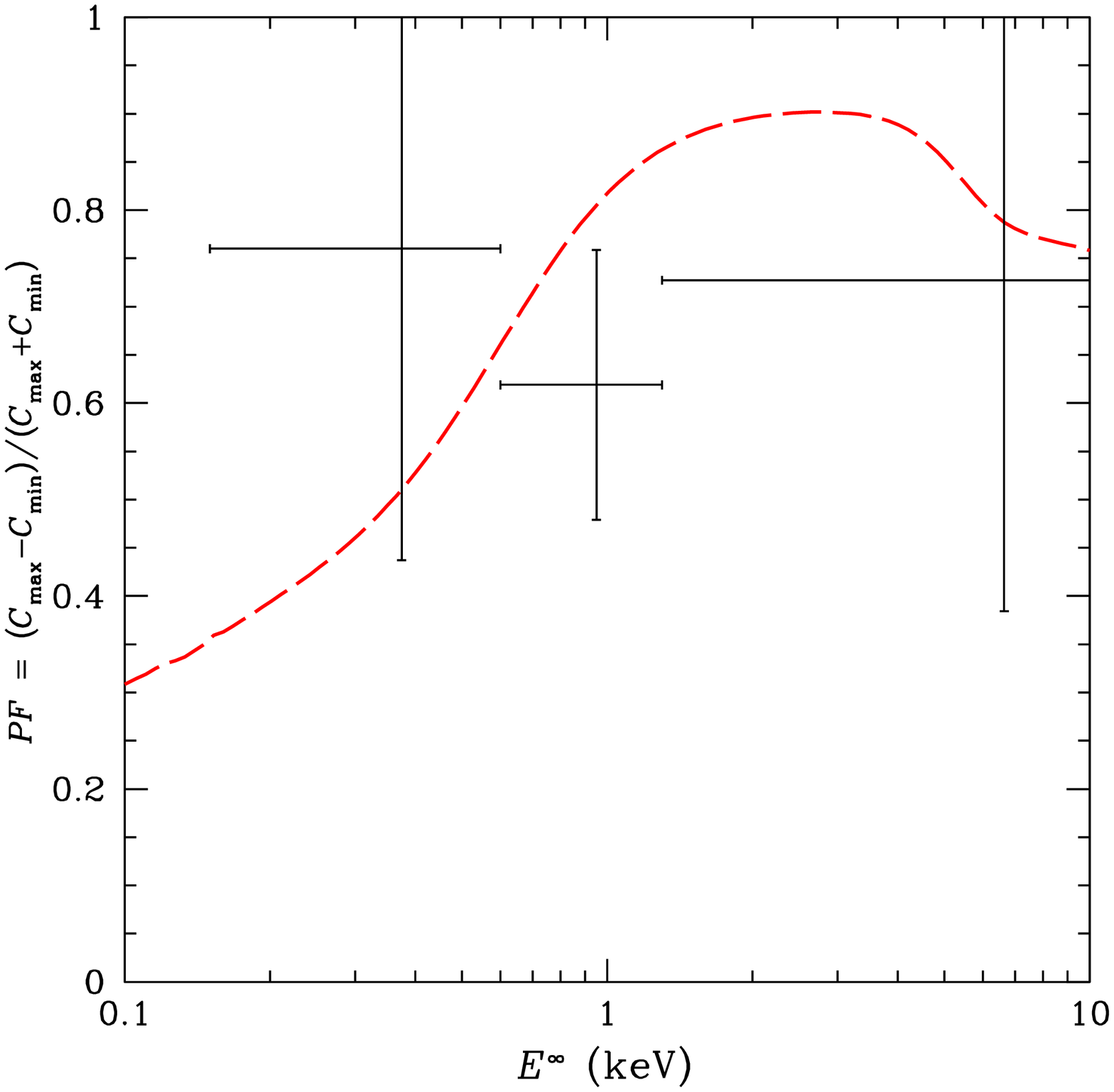}}}
\figcaption{Q mode X-ray pulse fraction as a function of energy for PSR~B0943+10.
The observed pulse fraction (crosses) are calculated from the pulse profiles
shown in Fig.~3 of M13 after subtracting the background
counts, and errors are $\approx 1\sigma$.
Dashed curve is the (redshifted) result of our atmosphere model (see Fig. 2 for model values).}
\end{inlinefigure}

\noindent ristics of atmosphere emission compared to blackbody emission
(see, e.g., Romani~1987; Shibanov et al.~1992).
The atmosphere emission radius (85~m) is much closer to that expected for a
centered dipole (140~m) than the blackbody emission radius.
Figures~2 and 3 show examples of our atmosphere model results for the pulse
profile and pulse fraction compared to the measurements of M13 
(errors are calculated from the square root of the number of
counts; c.f. Gehrels~1986).
We see that the atmosphere model shows promise, as long as the low-energy
pulse fraction is indeed $<100\%$.


\section{Light curves and the displaced dipole geometry}

In the previous section, we allowed for the low-energy pulse fraction
of the Q-mode emission to be relatively low. If future observations of
PSR~B0943+10 show
that the pulse fraction is high across all energy bands, then 
the ``canonical'' model of Sec.~3 cannot work.
We thus explore an alternative model for the emission geometry, in
which the magnetic dipole axis is displaced by a vector $\mathbf{s}$
from the center, such that $\theta_B$ is unchanged and thus the radio
emission geometry and polarization profile are not altered (see
Fig.~1), assuming that the radio emission originates beyond a few tens
of NS radii away from the surface.

Displaced magnetic dipoles have been invoked before to explain
various phenomena exhibited by pulsars. For example, Arons (2000)
demonstrated that the position of the pulsar death line can be altered
by displacing the magnetic dipole, and that to explain all observed active
pulsars displacements of as much as $(0.7-0.8)R_\star$ are needed. More recently,
Burnett \& Melatos (2014) showed that a displaced magnetic dipole may explain
some anomalies exhibited by the $Q-U$ phase portraits of some radio pulsars.

 \begin{inlinefigure}
\scalebox{0.55}{\rotatebox{0}{\includegraphics{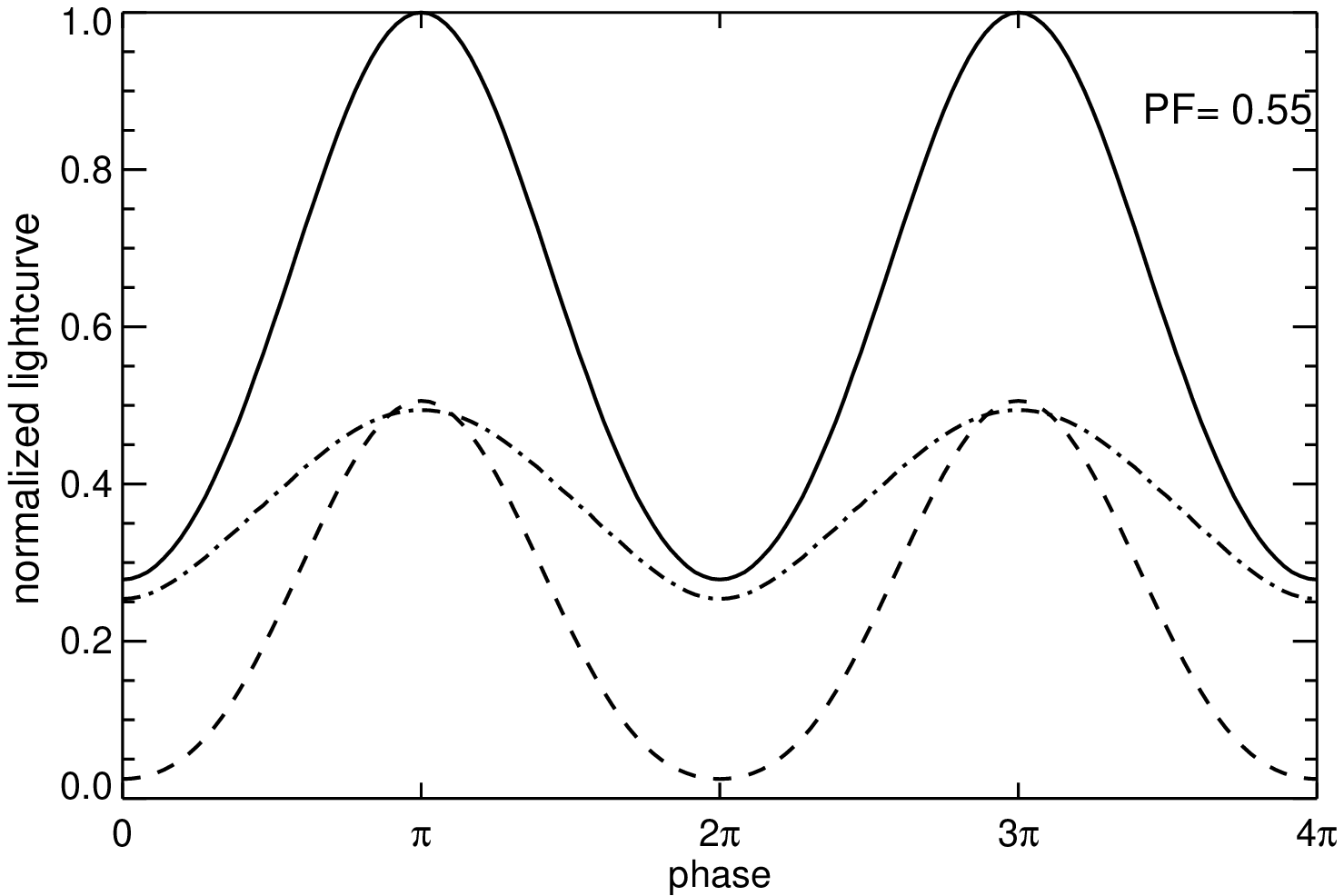}}}
\figcaption{Theoretical light curves for the displaced-dipole model of
  PSR~B0943+10 at $(0.2-12)$keV. See Fig. 1 for a schematic
  representation of the model; here, $\zeta=125^\circ$, and
  $(\beta,\theta_B)=(7^\circ,11^\circ)$. Each hot spot has
  $T_\mathrm{eff}=1.5\times10^6$K; the magnetic fields are
  $2\times10^{14}$G and $2.7\times 10^{12}$G for the near and far
  spots, respectively. The dot-dashed curve shows the B-mode 
  light curve due to the near spot with a pulsation of $31\%$. The dashed
  curve shows the emission from the far spot, visible only in
  the Q mode. The solid curve -- the sum of the near and far spots-- is the
  total Q mode light curve, with a pulsation of $55\%$.}
\end{inlinefigure}

We consider a $M=1.4M_{\odot}$, $R_\star=10$~km NS with a dipole
surface magnetic field $B_d\simeq 2\times 10^{12}$~G and propose the
following scenario: (i) \textit{All} or most of the
X-ray emission from PSR~B0943+10 is thermal in nature, which is, to
date, consistent with the analyses of the X-ray spectra carried out by
H13 and M13.  (ii) The magnetic dipole axis is transposed such that
both polar cap hot spots are potentially visible, as shown in Figure 1. The
larger (far) spot has a lower magnetic field, while the smaller (near) spot has a
larger magnetic field. In the Q mode, the large spot comes close to
being invisible (note that the strong light bending effects enable the
surface hot spots to be observed up to $135^\circ$ from the line of
sight), and thus produces a nearly-$100\%$ pulsation.  The observed Q
mode X-ray light curve is then composed of the nearly-constant
emission from the near spot, and the strongly pulsed emission from the
far spot.  (iii) The transition from the Q mode to the B mode involves a 
small
shift in the magnetosphere
or the dipole axis, such that the large spot -- which is already close
to being unseen -- becomes completely invisible, leaving only the
nearly-constant light curve of the small spot.

We denote by $\zeta$
the angle between the rotation axis and the large polar cap (see Fig.~1).  Quite
naturally, the only way to produce a thermal light curve with a
$100\%$ pulse fraction across \textit{all} energy bands is
through geometry: a hot spot must be placed such that it is close to
becoming invisible, in the range $\zeta\in(125^\circ,135^\circ)$. Then
the possible magnetic field axis orientations are constrained to a
conical shell of width $2\theta_B$ centered on the hot spot. However, 
since the observed peaks of X-ray and radio emissions from PSR~B0943+10 are
in phase (H13),
the magnetic axis must be oriented as shown in Fig. 1. It follows that the dipole must be
displaced by a distance $\mathrm{s}\approx (0.85-0.9)R_\star$, though its precise position,
as well as the effective temperatures of each hot spot, remain unknown. We do not attempt to explore
this parameter space. Rather, we present one test case to demonstrate
that, in principle, all the X-ray observations of PSR~B0943+10 can be
explained by this model.

 \begin{inlinefigure}
\scalebox{0.55}{\rotatebox{0}{\includegraphics{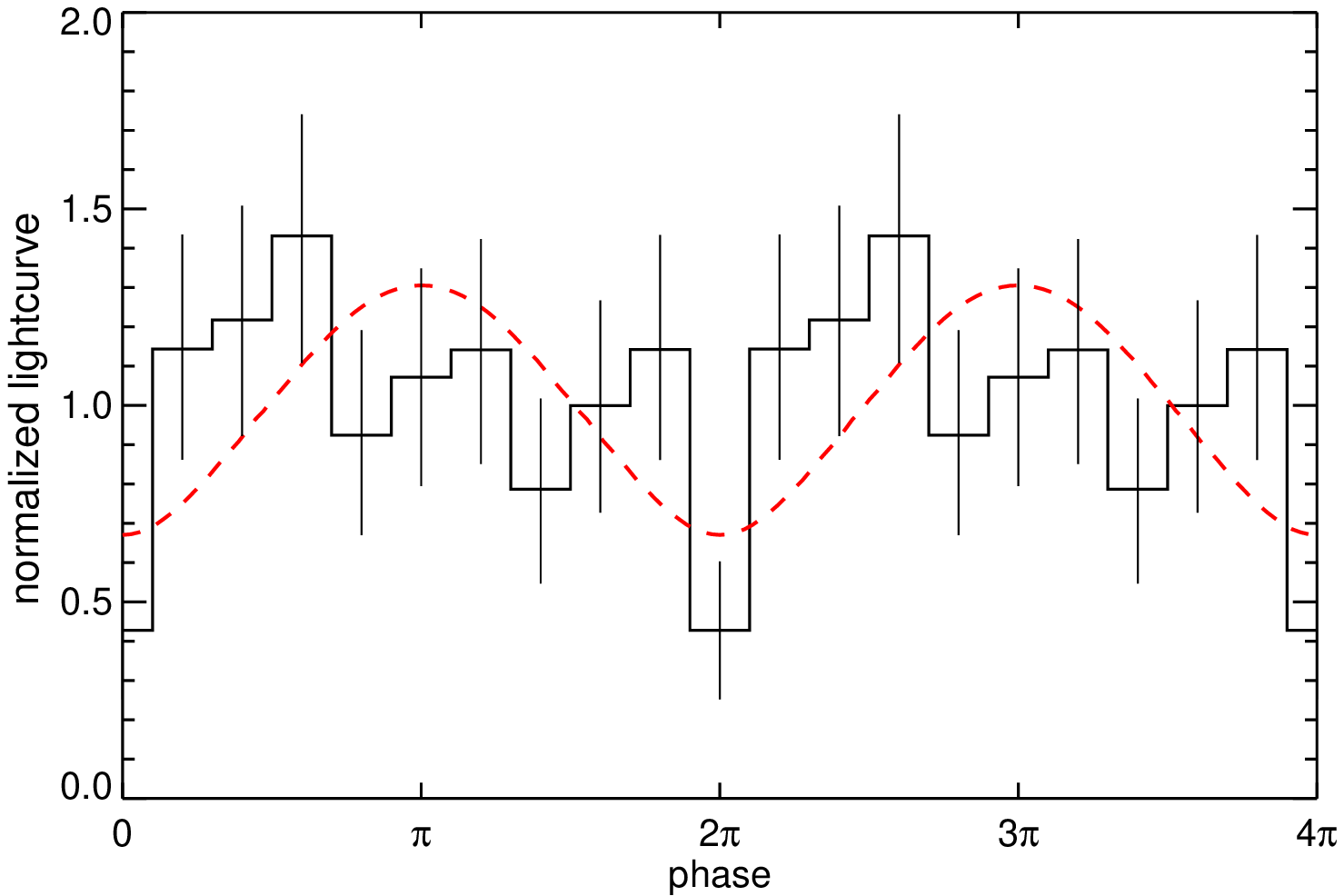}}}
\figcaption{Observed (solid histogram) and theoretical (red dashed) light
  curves for B mode X-ray emission of PSR~B0943+10 at
  $(0.2-12)$keV. The histogram is taken from Fig.~4 of M13. 
  Errors are $\approx 1\sigma$.
  Both the histogram and the theoretical curve have been
  normalized by their respective mean values. All parameters are same
  as those of Fig.~4, but we assume that in the B mode only the small
  spot is visible.}
\end{inlinefigure}

Naturally, the smaller ``near'' spot must be the source of the radio
emission. As discussed in section 2, the presence of drifting
subpulses in the radio emission likely implies the presence of a
vacuum gap at the hot spot surface, and a strong ($\go 10^{14}$G) surface magnetic field
is required to create such a gap. Therefore, we place the magnetic
dipole close to the surface of the near spot.

We use the method (``Temperature Template with Full Transport'', or
TTFT) developed in Shabaltas \& Lai (2012) to compute synthetic light
curves for our chosen hot spot configuration. The TTFT method uses
temperature profiles (as a function of depth) of magnetic
NS atmospheres to efficiently compute the observed surface thermal
radiation, taking into account the effects of magnetic atmosphere
opacities, beam pattern, vacuum polarization, and gravitational light
bending.  For this work, we have used realistic ionized hydrogen magnetic NS atmosphere
models (Ho \& Lai 2001,2003) to calibrate our temperature profiles for
the relevant effective temperature and surface magnetic field
values. 

As the blackbody fits of both the B mode and the Q mode spectra are
very similar, we fix the two hot spots to have equal temperature,
$T_\mathrm{eff}=1.5\times10^6$K. We fix the magnetic field strengths
to be $2\times10^{14}$G and $2.7\times 10^{12}$G for the small and
large spots, respectively; it follows that the hot spot radii must be
$\approx 20$m and $\approx 100$m. Figure 4 presents our results.

Satisfactory agreement with the Q mode observation is
achieved. However, the B mode light curve in this
particular case is not entirely constant; it has a pulse fraction of
$31\%$, coming mostly from the lower-energy part of the emission. We
note that this pulse fraction is within the $3\sigma$ upper limit (i.e., $\lesssim 50\%$) derived by
M13 for the B mode pulse fraction. Figure 5 shows
that the B mode theoretical lightcurve is consistent with the
most up-to-date observations. We speculate that a careful search in
the available phase space can reduce this pulsation, though it is not
likely to disappear entirely.


\section{Discussion and Conclusion}

The simultaneous radio and X-ray observations of the mode-switching
pulsar PSR~B0943+10 provide a unique opportunity for understanding this
class of objects. In this paper, we have examined several puzzles
presented by PSR~B0943+10. Our conclusions can be summarized
as follows.

We show that the large X-ray pulse fraction of the additional thermal component
observed in the radio Q mode can
be adequately reproduced
by using the canonical emission geometry of PSR~B0943+10, that is constrained
by radio observations, and taking into account beaming due to a strong surface
magnetic field. We find that the low-energy pulse fraction
produced by this model is small ($\sim 50\%$).

We also consider another, more extreme, magnetic field geometry that is consistent
with various observations, including the possibility that the extra
thermal component has high pulse fractions in all energy bands.
We show that by displacing the magnetic dipole by a 
large amount ($\sim 0.85-0.9 R_\star$) from the center of the star, both the B and the Q
mode X-ray observations can be explained as thermal emission coming
from the magnetic polar caps, one of which is small in size and has a large
surface magnetic field, thus being responsible for the radio emission,
while the other is large in size and has a lower magnetic field and acts as
the highly pulsed thermal component observed in the Q mode. 
We speculate that, when the pulsar is radio bright, the larger spot becomes invisible,
leaving only the smaller spot's emission to be observed. 

Overall, the models explored in this paper can be tested by future X-ray 
observations (e.g., the ``extreme'' model requires that most of the observed
X-rays are thermal in nature and predicts a $\sim 30\%$ pulse
fraction even in the B-mode). These will be valuable for constraining the
magnetosphere plasma density and the magnetic field strength and geometry 
of PSR~B0943+10, and will thereby further our understanding of the enigmatic behavior of
pulsar mode switching.

\medskip

W.C.G.H. appreciates the use of the computer facilities at KIPAC.
This work has been supported in part by NSF grants
AST-1008245, AST-1211061 and NASA grant NNX12AF85G.

\end{document}